\documentclass[12pt, a4paper]{iopart}

\usepackage[]{graphicx}  
\usepackage{wrapfig}
\usepackage{courier}
\usepackage{multirow}
\usepackage{cite}
\usepackage[rightcaption]{sidecap}

\usepackage{xcolor}

\begin{document}

\title{H-mode transition in the TJ-II stellarator plasmas}

\author{
T. Estrada, C. Hidalgo and the TJ-II team}

\address{Laboratorio Nacional de Fusi\'on, CIEMAT, 28040 Madrid, Spain}

\ead{teresa.estrada@ciemat.es}

\begin{abstract}
Since the first H-mode transitions were observed in TJ-II plasmas in 2008, an extensive experimental effort has been done aiming a better physics understanding of confinement transitions. In this paper, an overview of the main findings related to the L-H transition in TJ-II is presented including how the radial electric field is driven, which are the possible mechanisms for turbulence suppression, and what are the related temporal and spatial scales which can impact the transition.
The trigger of the L-H transition in TJ-II plasmas is found to be more correlated with the development of fluctuating $E\times B$ flows than with steady-state $E_r$ effects, pointing to the role played by zonal flows in mediating the transition. Experimental evidence supporting the predator-prey relationship between turbulence and flows as the basis for the L-H transition, found for the first time in TJ-II, reinforces this conclusion. Besides, the reduction in the turbulent transport at the transition is detected at the barrier region but also in a wider radial range with weak or even zero $E\times B$ flow shear, what points to other mechanisms beyond the turbulence suppression by local sheared flows.
\end{abstract}

\section{Introduction}

The High confinement mode (H-mode) regime has been extensively studied since its discovery in the ASDEX tokamak in 1982~\cite{Wagner:1982}.
The low to high confinement mode transition (L-H transition) is a bifurcation-type transition to improved confinement and reduced turbulence transport that develops spontaneously provided a power threshold is surpassed~\cite{Ryter:2002,Wagner:2007,Martin:2008}.
L-H transition is observed both in tokamaks and stellarators~\cite{Wagner:2006} and is characterized by an increase in plasma energy content and confinement time, consequence of a reduction in losses of particles and energy caused by the so-called transport barrier.

\maketitle

At the barrier the turbulence is strongly reduced and steep gradients in density, temperature and radial electric field develop. 
The earliest theories already ascribed a key role in triggering the L-H transition to the radial electric field and attributed the transition to an increase in the mean $E\times B$ flow and flow-shear, and the subsequent reduction in the level of edge turbulence~\cite{Itoh:1988, Biglari:1990,Carreras:1994,Connor:2000}. These models, however, did not account for the fast time scale associated to the transition, as later theories did revealing the crucial role of zonal flows (ZF) in triggering the transition~\cite{Kim:2003b,Malkov:2009,Miki:2012}.

From the experimental point of view, the key role of radial electric field in the development of the L-H transition has been demonstrated in several devices.  The question, however, is how the radial electric field is driven and what are the related temporal and spatial scales which can impact the transition and whether these scales end up in different routes to reach the transition conditions. 
Besides, the strongest reduction in the turbulent transport at the transition has been detected at the barrier region where the maximum $E\times B$ flow shear develops, but in some cases, a reduction in the turbulent transport is found in a wider radial range, with weak or even zero flow shear, what points to other mechanisms beyond the turbulence suppression by local sheared flows~\cite{Melnikov:2013}.

Since the first H-mode transitions were observed in TJ-II plasmas in 2008, an extensive experimental effort has been done to develop unique diagnostic capabilities in combination with advanced analysis tools, essential to provide further physics understanding of confinement transitions. In this paper, an overview of the main findings related to the L-H transition in TJ-II is presented.

\section{TJ-II properties, diagnostic capabilities and advanced analysis tools}

TJ-II is a heliac type stellarator with major radius $R = 1.5$ m, minor radius $<a> = 0.22$ m and magnetic field $B_0 < 1.2$ T. Plasmas are started up and heated by Electron Cyclotron Resonance Heating (ECRH) $2^{nd}$ harmonic using two gyrotrons at $53.2$ GHz with X-mode polarization. The maximum power per gyrotron is $300$ kW and the power density is about 10 W cm$^{-3}$~\cite{Fernandez:2001}. Under these heating conditions the plasma density has to be kept below the cut-off value of $1.7 \cdot 10^{19}$ m$^{-3}$. Higher density plasmas are achieved using Neutral Beam Injection (NBI) heating. Two injectors, one co- and one counter-, are in operation delivering a port-through power per injector up to 700 kW~\cite{Liniers:2017}. 
The flexibility of TJ-II offers the possibility to explore a wide rotational transform range in low, negative magnetic shear configurations. 

TJ-II is equipped with unique diagnostic capabilities able to measured turbulence and $E \times B$ flows in the entire plasma column and at different toroidal and poloidal positions, thus enabling the detection of ZFs.
The identification of  ZF structure, symmetric in the poloidal and toroidal direction and with a finite radial wavelength, requires direct measurement of flows at different radial, poloidal and/or toroidal positions, with good temporal and spatial resolution.  
Experimentally, ZFs are identified by applying the so-called long range correlation (LRC) analysis technique. This analysis computes the cross-correlation between two distant flow measurements taken on the same flux surface but not connected by a magnetic field line. A high correlation level and a cross-phase between the signals close to zero are signatures of a ZF.
Besides, to identify and characterize the interaction between flows and turbulence, both quantities should be measured simultaneously. Only a limited number of plasma diagnostic systems exist appropriate to measure both turbulence and $E\times B$ flows simultaneously and with good enough spatial and temporal resolution.
At TJ-II, this diagnostic set includes Langmuir probes~\cite{Pedrosa:2008,Hidalgo:2009}, Heavy Ion Beam Probe systems~\cite{Melnikov:2007,Melnikov:2015} and Doppler reflectometry~\cite{Happel:2009}.

Langmuir Probes (LP) are small-sized electrodes, which, when inserted in the plasma periphery, allow us to measure a combination of plasma parameters by measuring the current / voltage characteristic. Multiple pin probes are often used to determine radial profiles of the plasma parameters and to characterize the spatial structure of their fluctuations. Besides, due to their simplicity, dual Langmuir probe systems that are poloidally and/or toroidally separated are in operation in a number of fusion devices, as in TJ-II, allowing the identification of zonal flows at the plasma periphery.
Heavy Ion Beam Probe (HIBP) diagnostic allows the measurement of the plasma electric potential, the electron density and their fluctuations, well inside the plasma column where material probes cannot be inserted.  
The complexity of this diagnostic is considerable and increases as the magnetic field used to confine the plasma increases. Nevertheless, dual HIBP systems have been implemented in some experiments, the first one in CHS~\cite{Fujisawa:2004} and more recently also in TJ-II~\cite{Melnikov:2015}.
Doppler reflectometry (DR) is used to measure the density turbulence and its perpendicular rotation velocity at different turbulence scales. Its capability to measure simultaneously turbulence and flows with good spectral, spatial and temporal resolution makes DR a very attractive diagnostic presently used in several fusion devices. At TJ-II, a two channel, optimized system able to probe different turbulence scales at two separate poloidal regions is in operation since 2009~\cite{Happel:2009}.

The simulation and experimental measurement of plasma bifurcations generates vast datasets that need to be properly processed before they can lead to any actual insight on the physical mechanisms involved.  Standard turbulence analyses involve the calculation of the fluctuation amplitude, the probability distribution function and power spectra of the experimental time series. Normal mode oscillations can be recovered using linear data-driven methods such as the Biorthogonal Decomposition while nonlinear analysis can be done by computing the bicoherence~\cite{Milligen:1995b}. Going beyond these techniques, causality analysis based on the Transfer Entropy allows disentangling the causal relation between fluctuating fields and to study the propagation of fluctuations~\cite{Grenfell:2020}.

\section{H-mode transition}

H-mode transitions in TJ-II~\cite{Sanchez:2009} are achieved under NBI heating conditions when operating in lithium coated wall conditions~\cite{Tabares:2008}.  At the transition, the plasma density and plasma energy content increase and a steep edge density gradient develops with an increase in the mean $E\times B$ flow shear and a pronounced reduction in the plasma turbulence~\cite{Estrada:2009}. This phenomenology is similar to that observed in other helical devices and also in tokamaks, confirming as a generic regime of toroidal confinement. 
There are, however, some peculiarities observed in stellarators, like the strong dependence on the magnetic topology~\cite{Sano:2005,Wagner:2006,Hirsch:2008,Estrada:2009}, that may provide additional information to the L-H transition physics.
Besides, while the edge density gradient increases substantially at the transition, the temperature profiles remain almost unaffected in stellarators, except in the fully developed quiescent H-mode in W7-AS~\cite{Hirsch:2008}. The absence of a clear barrier in temperature profiles indicates a decoupled energy and particle transport as observed in the so-called I-mode in tokamaks~\cite{Hubbard:2011,Ryter:2016,Happel:2016,Hubbard:2016}.  Transport channel decoupling could be driven by any mechanism that leads to a modification of the cross-phase between fluctuating fields~\cite{Hidalgo:2016,Manz:2021}. Contrary to the H-mode in stellarators, in the I-mode in tokamaks the edge transport barrier develops in the temperature profiles and not in the density profile which remains similar to that in L-mode. 
Finally, neoclassical transport enforces a negative $E_r$ at the plasma edge with a substantial sheared flow already in L-mode, what may constitute a biasing precondition for the L-H transition in stellarators.

\subsection{$E\times B$ flow and turbulence at the transition}

\begin{figure}
\center
\includegraphics[width=0.5\columnwidth,trim= 0 0 0 0]{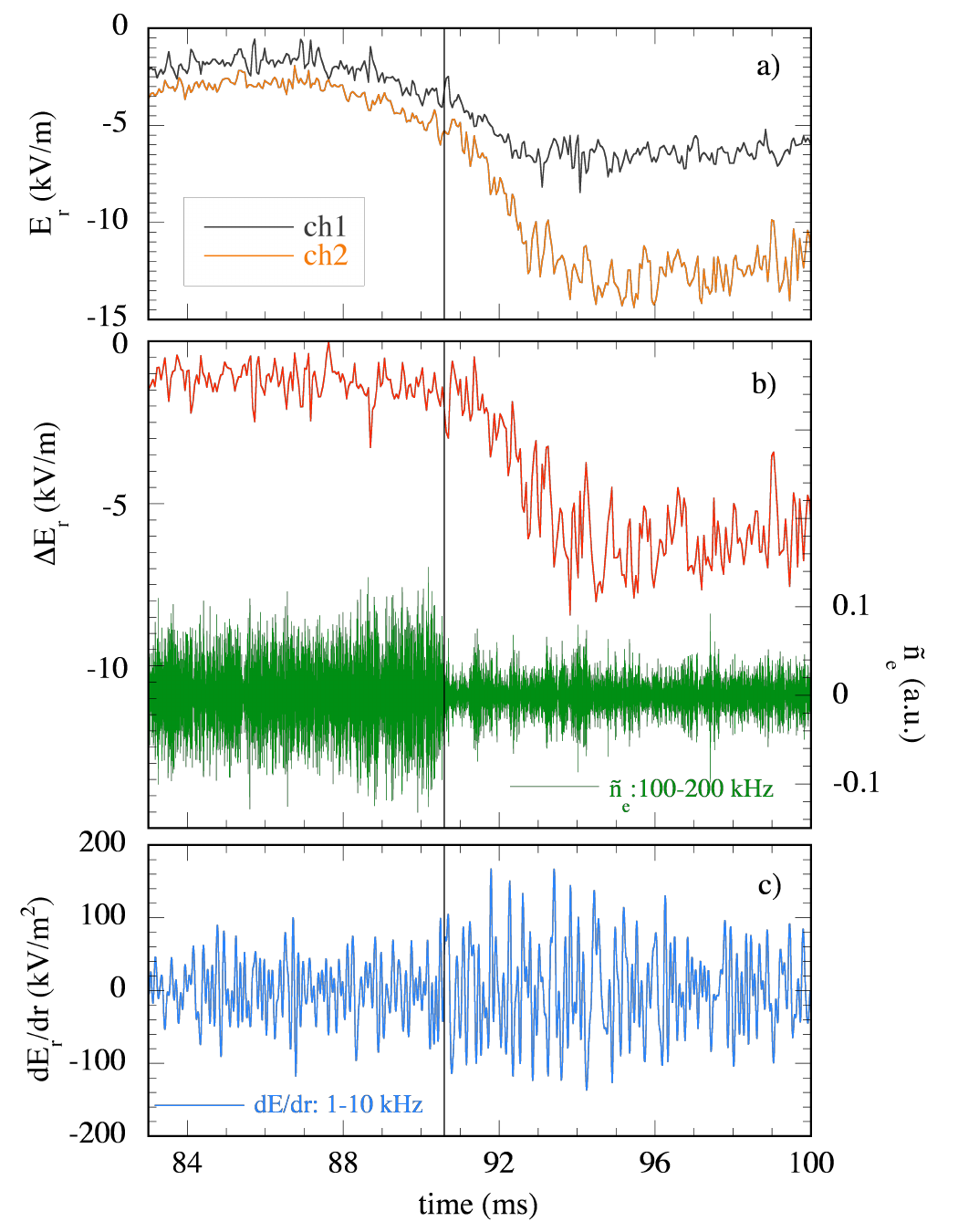}
\caption{Turbulence and $E\times B$ flow shear measured using Doppler reflectometry during the L-H transition in the TJ-II stellarator. (a) Time evolution of $E_r$ measured at two adjacent radial positions right at the $E\times B$ flow shear location: ch1 at $\rho$ = 0.86 and ch2 at  $\rho$ = 0.82. (b) $E_r$ shear (in red) and density fluctuations (in green), and (c) the low frequency components (1-10 kHz) of the fluctuating $E_r$ shear. The vertical line indicates the time of the L-H transition. Figure reproduced with permission from~\cite{Estrada:2009}. Copyright 2009 IOP Publishing.}
\label{fig_sim_1}
\end{figure}

In stellarators, neoclassical transport enforces a substantial negative $E_r$ already in L-mode. 
At the L-H transition, $E_r$ becomes more negative and a pronounced $E\times B$ flow shear develops together with an abrupt reduction in plasma turbulence close to the radial position of maximum $E\times B$ flow shear.
At the H-mode, the well in $E_r$ deepens by a factor of up to 2 more than expected from the increase in diamagnetic term alone~\cite{Hirsch:2008,Estrada:2009}.
As in tokamaks, the experimental observations indicate a close relation in the turbulence and flow evolution, supporting the picture of turbulence suppression by sheared flows~\cite{Itoh:1988, Biglari:1990,Carreras:1994,Connor:2000} as a basic element of the transition physics. Most observations, however, do not account for the fast time scale associated to the transition to H-mode and as a consequence, the question of causality can be hardly addressed.
To properly describe the dynamics of the transition, very fast and radially localized measurements of the relevant quantities are needed. Doppler Reflectometry (DR) is one of the tools suitable for these type of measurements as it allows to study the dynamics of the $E_r$, $E_r$ shear and density fluctuations with very good temporal and spatial resolution~\cite{Hirsch:2001}.
At TJ-II, DR measurements showed that the reduction in density fluctuations at the transition precedes the increase in the mean $E\times B$ flow shear.
This can be seen in figure \ref{fig_sim_1} where the time evolution of $E_r$ measured at two adjacent radial positions right at the $E\times B$ flow shear location is shown in figure \ref{fig_sim_1}.a, and the $E_r$ shear (in red) and density fluctuations (in green) are displayed in figure \ref{fig_sim_1}.b.
This result resembles that found in JET~\cite{Andrew:2008} where large values of the $E_r$ shear, measured during the H-mode, do not appear to be necessary for the L-H transition. More recent measurements performed using DR in JET shows a shallow $E_r$ at the transition pointing to similar conclusion~\cite{Silva:2021}.
These results appeared to be in contradiction to numerous experimental observations supporting the paradigm of turbulence suppression by $E_r$ shear to explain the transitions.
However, a detailed analysis of the signals measured at TJ-II revealed an increase in the fluctuations of $E_r$  and $E_r$ shear within the frequency range 1-10 kHz just at the transition~\cite{Estrada:2009}. 
The intensity of the low frequency components of the fluctuating $E_r$ shear is shown in figure \ref{fig_sim_1}.c. 
The increase in the low frequency $E_r$ fluctuations and the reduction in the density fluctuations are synchronous and take place before the mean $E\times B$ flow shear development.

These results indicate that the trigger of the L-H transition is more correlated with the development of fluctuating $E\times B$ flows than steady-state flow shear effects, what may be an indicator, albeit insufficient, of the leading role played by zonal flows.
Additional experimental measurements provided by Langmuir probes reinforced this conclusion.
The fluctuating $E\times B$ flows arising at the transition were detected using Langmuir probe arrays located at the plasma periphery at two distant toroidal positions~\cite{Hidalgo:2009}. 
Long range correlation (LRC) analysis applied to these flow measurements showed a high correlation level and a cross-phase between the signals close to zero when the L-H transition happens. Same analysis applied to density fluctuations showed that these are always dominated by short-range scales.

\begin{figure}
\center
\includegraphics[width=0.5\columnwidth,trim= 0 0 0 0]{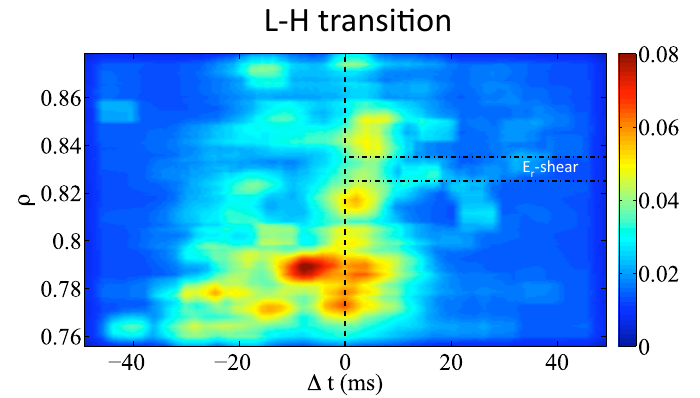}
\caption{Radially resolved auto-bicoherence measured during the L-H transition in the TJ-II stellarator. The vertical dashed line indicates the L-H transition time defined as in figure 1. The $E_r$ shear position is marked in the figure. Figure reproduced with permission from~\cite{Milligen:2013}. Copyright 2013 IAEA, Vienna.}
\label{fig_sim_2}
\end{figure}

Further signatures of H-mode transition triggered by zonal flows were found applying non-linear analysis~\cite{Milligen:2013}.
Nonlinear analysis, although an indirect approach, can be used as an indicator of increased zonal flow generation.
The degree of nonlinear three-wave coupling can be measured by computing the bicoherence. The relation between the bicoherence and the zonal flow growth, involving an interaction between high frequencies associated with drift wave turbulence and low frequencies associated with the zonal flow, was shown in~\cite{Diamond:2000}. Thus, an increase in the three-wave coupling, i.e., in the bicoherence, is required if the zonal flow is generated by the turbulent Reynolds stress.
Spatiotemporal resolved bicoherence analysis was applied to study the nonlinear coupling between turbulence and $E\times B$ flows measured by DR during the L-H transition~\cite{Milligen:2013}. The analysis showed a pronounced nonlinear coupling between high-frequency turbulence and low-frequency fluctuations, as shown in figure \ref{fig_sim_2}. The increase in the non-linear coupling is observed a few milliseconds before the L-H transition time, at a given radial position (inward from the $E\times B$ shear layer in H-mode). Then, the coupling expands in radial space through the time of the transition.
These spatiotemporal sequence of events agrees qualitatively with the development of zonal flows at the L-H transition shown in the 1D spatiotemporal model reported in~\cite{Miki:2012}. In this model the zonal flow initially peaks in a narrow radial region right before the L-H transition, then expands in radius through the time of the transition, and shortly afterwards disappears.
These results point to the role played by zonal flows in mediating the transition into the H-mode.

As already mentioned, an abrupt reduction in plasma turbulence and associated turbulent particle flux is detected in the radial region of maximum $E\times B$ flow shear at the L-H transition. This reduction, however, is not necessarily restricted to the region of maximum flow shear.
Reduction in the turbulent transport has been also reported, for example, at the curvature region of the $E_r$-well structure in H-mode~\cite{Kobayashi:2017}. In these experiments, the density turbulence is substantially suppressed at both the inner and outer $E_r$ shear regions while the variation is only moderate in the curvature region. The phase difference between the density and the potential fluctuations, however, approaches zero in the inner shear and curvature regions, resulting in an overall turbulent particle flux suppression observed in a wide radial range.
HIBP diagnostic allows these type of measurements not only in the plasma periphery but also in the core region.
In TJ-II, HIBP measurements have shown a reduction in the plasma turbulence and associated turbulent flux not only in the plasma edge region but also in the plasma core at $\rho \sim 0.5$~\cite{Melnikov:2013}. 
The changes in the turbulent flux are detected in the two plasma regions simultaneously within a time resolution of $\sim$1 ms.  
A change in the cross-phase between the density and the potential fluctuations has been also detected using Langmiur probes at the plasma edge as the plasma transits from electron to ion root confinement regime in TJ-II~\cite{Milligen:2021}.
In all cases, the cross-phase and the turbulence evolve in different time scales what may be an indication of the existence of different underlying mechanisms~\cite{Kobayashi:2020}.

The turbulence reduction at the plasma core region, where the $E\times B$ flow shear is not substantially high, could be explained in terms of the combined effect of the $E\times B$ flow shear at the edge and the radial spreading of turbulence: as the edge flow shearing increases, it reduces the turbulence spreading from the plasma edge to the core.
Signatures of such effect have been found in the H-L back-transition in TJ-II~\cite{Estrada:2011}. The spatiotemporal evolution of the density turbulence and $E\times B$ flow were measured by DR as the plasma approached the H-L back-transition. The results are explained as a combined effect of the $E_r$ shear reduction and the radial spreading of turbulence from the plasma core to the edge barrier and suggest the following scenario:  The radial spreading of the turbulence, braked during the H-mode by the high $E_r$ shear, becomes noticeable as the shear declines and produces a gradual increase in the turbulence level at the innermost radial positions, reaching the $E_r$ shear location right before the H-L back-transition. The consequence is a gradual retreat of the transport barrier. More recently, the key role of the edge  $E_r$ shear in controlling the turbulence spreading from the edge to the Scrape-Off Layer (SOL) has been also identified in TJ-II~\cite{Grenfell:2020}.  These results resemble the simulation studies reported in~\cite{Wang:2007}, where the key quantity to the control of turbulence spreading was found to be the $E_r$ shearing rate.

The so-called isotope effect is observed in tokamaks as an increase of the energy confinement time with the ion mass. An important aspect related to this effect is the impact it has on the L-H transition power threshold~\cite{Righi:1999,Ryter:2013}.  A reduction of the L-H power threshold by about 50$\%$ is found when using Deuterium or Helium instead of Hydrogen. Experiments performed in stellarators, however, show that the ion mass has in general a much tinier impact on the confinement time.  
The impact of the ion mass on the L-H transition has been studied in TJ-II plasmas. Pure Hydrogen vs. Deuterium dominated (up to 70\%) plasmas did not show any noteworthy difference when comparing $E_r$ profiles, amplitude of ZFs or plasma turbulence reduction during the L-H transition~\cite{Losada:2018}. Interestingly, the radial width of the ZF structure is affected by the ion mass, being about 1.5 times larger in D than in H plasmas~\cite{Losada:2021}.

 \subsection{Intermediate or LCO phase}
 
\begin{figure}
\center
\includegraphics[width=0.5\columnwidth,trim= 0 0 0 0]{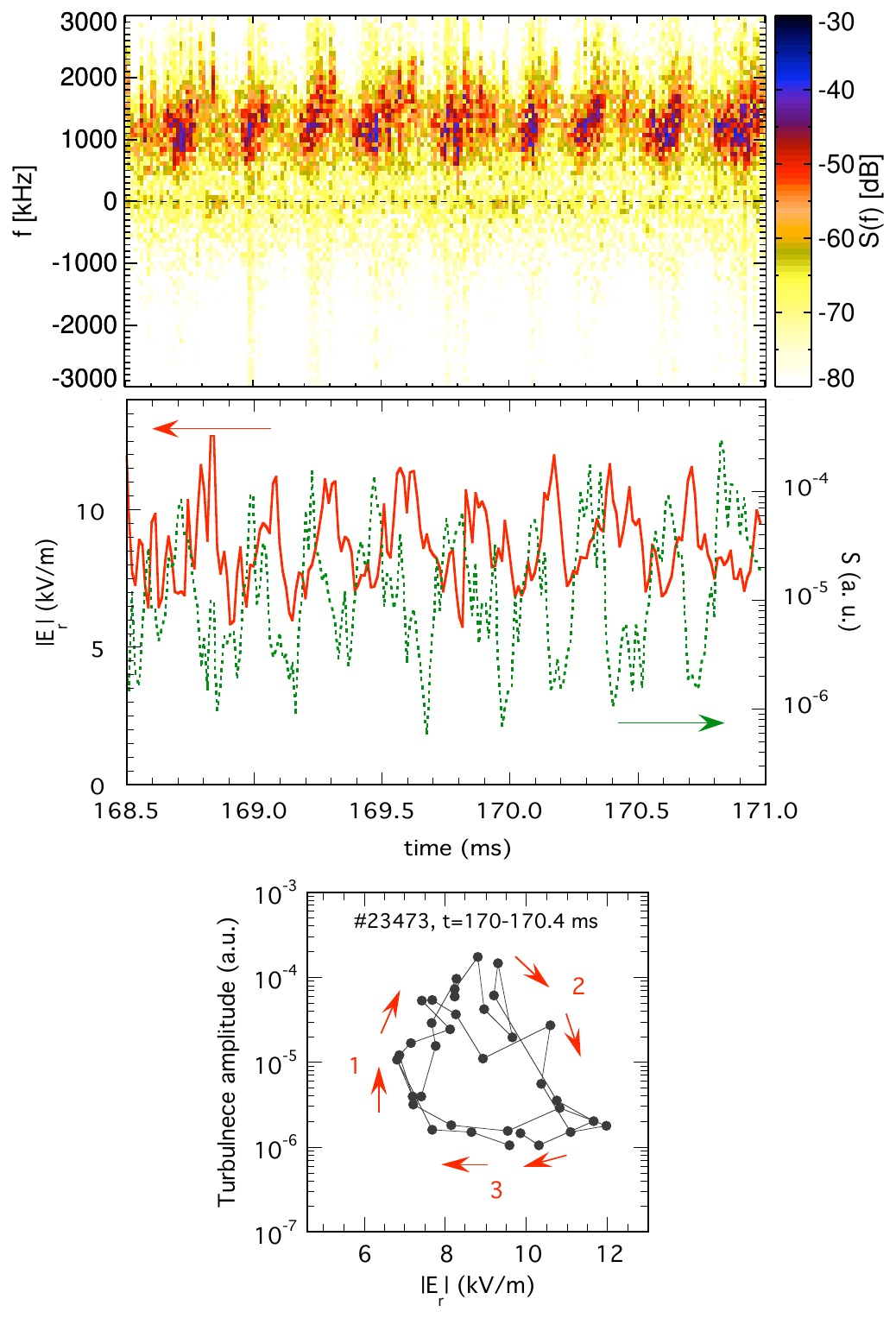}
\caption{Time evolution of $E\times B$ flow and density fluctuations measured by Doppler reflectometry during the I-phase in the TJ-II stellarator. Top: Spectrogram of the Doppler reflectometer signal measured at $\rho \sim 0.8$, in a magnetic configuration with $\iota/2\pi = 1.53$. The color code reflects the density fluctuation level and the frequency of the Doppler peak gives $E_r$. Middle: Time evolution of $E_r$ (red solid line) and density fluctuation level (green broken line) obtained from the spectrogram. Bottom: Relation between density fluctuation level and $E_r$ in phase space during the I-phase; only two of the cycles are displayed. The time interval between consecutive points is 12.8 $\mu s$. Figure reproduced with permission from~\cite{Estrada:2010c}. Copyright 2010 Europhysics Letters Association.}
\label{fig_sim_3}
\end{figure}

 The results described in the previous section point to the  possible role that zonal flows can play in mediating the transition into the H-mode. In this section, the so-called Intermediate phase (I-phase) or Limit Cycle Oscillation (LCO) phase is discussed.
 Bifurcation theory models based on the coupling between turbulence and flows describes the L-H transition passing through an intermediate, oscillatory transient stage~\cite{Kim:2003b}. 
In these models, zonal flows trigger the transition by regulating the turbulence until the mean shear flow is high enough to suppress turbulence effectively, which in turn subsequently impedes the zonal flow generation. 
 Because of the self-regulation between turbulence and flows, the transition is marked by an oscillatory behavior with a characteristic predator-prey relationship between turbulence and zonal flows. 
 The first experimental evidence supporting the predator-prey relationship between turbulence and flows as the basis for the L-H transition was found in TJ-II~\cite{Estrada:2010c}. Close to the transition threshold conditions, gradual transitions are achieved showing an intermediate, oscillatory transient phase. The temporal dynamics of turbulence and flows, as measured by DR, shows pronounced oscillations right inside the $E_r$ shear layer, with a coupling between them which reveals a characteristic predator-prey relationship: a periodic behavior with the flow (predator) following the turbulence (prey) with $90^\circ$ phase difference.
 This result is shown in figure  \ref{fig_sim_3}. The spectrogram of the DR signals is displayed in the top panel of the figure. The contour color map reflects the amplitude of the Doppler peak which is proportional to the density fluctuation level, while the frequency of the Doppler peak gives $E_r$. These magnitudes, obtained by fitting a Gaussian function to the spectra, are also shown in the figure (middle panel) together with the relation between them in phase space showing the limit-cycle behaviour (bottom panel). 
Besides, the spatiotemporal evolution of the turbulence-flow oscillation pattern was also measured, showing both radial outward and inward propagation velocities of the turbulence-flow front~\cite{Estrada:2011b}. The results indicate that the edge shear flow linked to the L-H transition can behave either as a slowing-down, damping mechanism of outward propagating turbulent-flow oscillating structures, or as a source of inward propagating turbulence-flow events. The spatiotemporal evolution of the phenomenon detected at TJ-II triggered the extension of the theoretical models to study the spatial propagation properties~\cite{Miki:2012,Miki:2013}.

Further experimental evidence of the predator-prey relationship during the LCO phase has been also found in other devices when operating close to the L-H transition threshold conditions~\cite{Xu_GS:2011,Schmitz:2012,Kobayashi:2013,Tynan:2013}. 
In particular, the measurements carried out in the DIII-D tokamak using a multichannel DR allowed a direct characterization of the spatiotemporal structure of the LCO finding radial inward propagation velocities starting at the separatrix position. Besides, at the onset of the LCO, the $E\times B$ flow oscillations lag the density fluctuation level by $90^\circ$, as found in TJ-II, transitioning to anti-correlation ($180^\circ$) when the equilibrium shear dominates the turbulence-driven flow shear due to the increasing edge pressure gradient, ending up in a transition to a sustained H-mode. 
In the tokamak HL-2A~\cite{Cheng:2013}, two types of limit-cycle oscillation were found  showing opposite temporal ordering. In the first type of limit cycle, the turbulence intensity grows first, followed by the increase of the $E\times B$ flow, similar to previously reported observations. This is the standard predator-prey model where the turbulence increase leads the zonal flow generation which suppresses the former. In the second type of limit cycle, the $E\times B$ flow grows first, causing the reduction of the fluctuations, what points to the pressure gradient as a candidate for maintaining the oscillations and eventually inducing the I-H transition. This second limit cycle type may be explained by the bifurcation model of the L-H transition reported in~\cite{Itoh:1988}.
Posterior experiments carried out in TJ-II also identified the two types of limit-cycle oscillation showing that the change in the temporal ordering is not related to an intrinsic change in the nature of the LCO~\cite{Estrada:2015}. LCO propagating inwards results in a clockwise (CW) LCO, as in the standard predator-prey model. On the contrary, LCO propagating outwards results in a counterclockwise (CCW) LCO when the local $E\times B$ flow is considered. However, if the local $E\times B$ flow shear is considered for the LCO representation, the CW temporal ordering is recovered.  In the two LCO types, the turbulence increase leads the process and produces an increase in the $E\times B$ flow shear. 
These results provide strong experimental evidence that zonal flow and equilibrium shear flow are instrumental in the L-H transition, supporting the predator-prey type models.

  \subsection{Scale-resolved turbulence studies}

\begin{figure}
\center
\includegraphics[width=0.5\columnwidth,trim= 0 0 0 0]{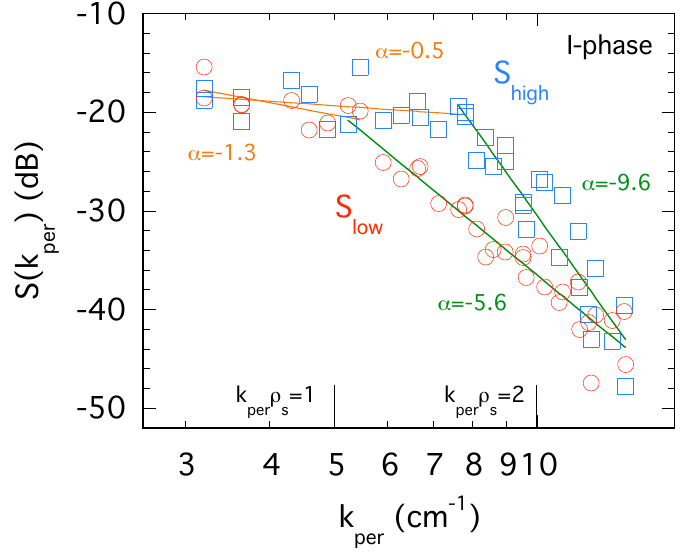}
\caption{Density fluctuation wavenumber spectra measured during the I-phase by Doppler reflectometry in the TJ-II stellarator. The extreme values of the turbulence level measured during the oscillatory phase are represented: maxima (in blue) and minima (in red). The spectral indexes are shown for each spectrum and the normalized wave-numbers $k_\perp \rho_s = 1$ $\&$ $2$ are marked in the figure. Figure reproduced with permission from~\cite{Estrada:2012b}. Copyright 2012 IOP Publishing.}
\label{fig_sim_4}
\end{figure}

Scale-resolved turbulence studies can help in the identification of the underlying processes responsible for turbulence suppression at the transition.
Scale-resolved measurements of the turbulence reduction in H-mode plasmas has been addressed in TJ-II~\cite{Happel:2011}.
To that end, the density turbulence wavenumber spectrum was measured using DR in L- and H-mode plasmas.
In both scenarios, two wavenumber regions, with different power laws and a well defined knee, can be identified where the turbulence decays slowly at at low $k_\perp$ and shows a faster fall off at higher $k_\perp$.
In H-mode, the turbulence reduction is observed to happen preferentially on intermediate scales, close to the $E_r$ shear layer (reduction by about one order of magnitude in comparison to L-mode). The preferential turbulence reduction takes place on scales $k_\perp$ $\sim 7-11$ cm$^{-1}$ ($k_\perp \rho_s$ $\sim 1.0-1.7$). 
At smaller radii, where $E_r$ is strong but its shear is moderate, the turbulence is reduced as well, but not as much as in the $E_r$ shear layer.
Similarly, the spectral structure of the turbulence-flow interaction has been measured during the LCO phase allowing the identification of the relevant turbulence scales involved in the turbulence-flow predator-prey process~\cite{Estrada:2012b}.
The density fluctuation wavenumber spectra measured during the intermediate oscillatory phase is shown in figure  \ref{fig_sim_4}. The extreme values of the turbulence level measured during the I-phase are displayed, the maxima in blue and the minima in red. 
The measurements indicate that intermediate turbulence scales within the range $k_\perp \sim 6-11$ cm$^{-1}$ dominate the process.  
In these experiments, a very slight dependence of the  flow oscillation amplitude on the turbulence scale is found indicating that all turbulence scales follow the flow oscillations although the scales involved in the energy transfer of the predator-prey process are preferentially the intermediate ones.
 The intermediate turbulence wavenumber range being the dominant player in the zonal flow generation by Reynolds stress has been identified as well both, experimentally~\cite{Manz:2009, Stroth:2011} and in simulations~\cite{Scott:2005}. The preferential reduction of intermediate turbulence scales can be explained within the framework of vortex straining and subsequent energy transfer through Reynolds stress into zonal flows, as suggested in~\cite{Manz:2009, Stroth:2011}.

The dependences of the zonal flow on possible driving and damping mechanisms have been studied at TJ-II using probes. The results show that the turbulence drive, the Reynolds stress acceleration, was able to provide the necessary perpendicular acceleration, however a causal relation could not be established. In addition, the calculated neoclassical viscosity and damping times were comparable to the observed zonal flow relaxation times~\cite{Alonso:2012,Gerru:2019}.
These results indicate that both turbulent and neoclassical mechanisms, and their mutual interactions, should be considered to achieve a full understanding of the dependences of edge zonal flows.

  \subsection{Impact of magnetic topology}

In low magnetic shear helical devices, the magnetic topology has a sensitive influence on the H-mode realization and quality.
 In W7-AS, specific magnetic configurations for accessing the transition were identified~\cite{Hirsch:2008}; quiescent H-mode could be reached in configurations with selected rotational transform windows only. The boundaries of these windows were located near natural magnetic islands.  Similarly, H-mode quality in Heliotron-J strongly depends on its magnetic configuration~\cite{Sano:2005}. A possible mechanism invoked to explain the experimental observations is the minimization of the neoclassical poloidal viscosity in the relevant configuration windows~\cite{Wobig:2000}.
In TJ-II, dedicated experiments scanning the magnetic configuration have shown a clear dependence of the H-mode realization and quality on the edge rotational transform~\cite{Estrada:2010a}. Both, the relative increase in the confinement enhancement factor over the L-mode value and the edge $E_r$ well reach the highest values in magnetic configurations with a low order rational surface at the plasma edge. 
 Moreover, magnetic configurations with the rational 5/3 close to the plasma edge show that the L-H transition requires certain plasma current that depends on the magnetic configuration, which indicates a preferential radial position for the rational to ease the transition.
These observations are interpreted in terms of local changes in $E_r$ induced by the rational surface that could facilitate or even trigger the L-H transition when the spin-up of the $E\times B$ flow takes place at the relevant edge layer.
This interpretation is supported by experimental results showing the impact of the magnetic islands on flows and turbulence in TJ-II~\cite{Bondarenko:2010,Estrada:2016} as well as in other devices~\cite{Ida:2001,Zhao:2015,Bardoczi:2016,Choi:2017,Estrada:2021b} and in numerical simulations~\cite{Banon_Navarro:2017}.

  \subsection{Impact of fast particles}

 The particle loss associated to the energetic particles driven Alfv\'en Eigenmodes (AE) can deteriorate the confinement but may have also an indirect, positive effect on turbulence and confinement.
 The coexistence and interplay of fast ions and instabilities is an active area of research.
Recent theoretical studies have shown that fast ions can strongly suppress turbulent transport; the key mechanism for this suppression being the resonant wave-particle interaction between fast ions and turbulence-driven microinstabilities~\cite{DiSiena:2020}. Gyrokinetic simulations have linked a substantial turbulence stabilization to the presence of fast particles in strong electromagnetic regimes. 
Non-linear excitations of zonal structures by fast particle driven modes have been predicted~\cite{Chen:2012,Qiu:2019,Mazzi:2020} but complete experimental validation is still missing. The plasma scenarios where fast particle driven zonal structures would affect the nonlinear dynamics of AEs and turbulent transport is at present an active area of research.

In TJ-II, AE activity driven by NBI-produced fast ions, with frequency and properties dependent on the magnetic configuration, has been studied~\cite{Jimenez:2007} and in some particular cases, a correlation between AE activity and edge ion losses has been observed. The particle loss associated to the energetic particles driven AE can deteriorate the confinement but may have also an indirect, positive effect on confinement by inducing local changes in $E_r$ with impact on the turbulence regulation by $E\times B$ flows. 
 The relation between zonal structures and AE have been experimentally studied using HIBP~\cite{Hidalgo:2022}. LRCs are detected both at the AE and at the zonal flow frequencies, but so far no causal relationship could be established between the amplitude of zonal flows and AEs. 
Besides, TJ-II experiments have shown that AE can have an impact on the transitions to H-mode. The results have been interpreted in terms of a geodesic acoustic mode generated by the non-linear mode coupling of AE that evolves towards a low frequency flow, as precursor of the transition~\cite{Estrada:2015}.

\section{Conclusions}

The L-H transition studies carried out in TJ-II during the last decade have permitted a rather complete characterization of the transition process revealing both, common features to those observed in tokamaks and other helical devices, and also some particular aspects that may provide valuable information to the L-H transition physics. 

The trigger of the L-H transition in TJ-II plasmas is found to be more correlated with the development of fluctuating $E\times B$ flows than with steady-state $E_r$ effects, pointing to the role played by zonal flows in mediating the transition. Experimental evidence supporting the predator-prey relationship between turbulence and flows as the basis for the L-H transition was found for the first time in TJ-II. Since them, different aspects of the LCO phase have been characterized triggering the extension of the theoretical predator-prey type models.
For recent reviews on the turbulence-flow interaction measured at the transition in different toroidal devices, and on the progress towards physics-based transition models, the reader is referred to~\cite{Schmitz:2017,Bourdelle:2020}

The observed impact of the magnetic topology on the transition at TJ-II has been interpreted in terms of a local spin-up of the $E \times B$ flow induced by a rational surface when taking place at the relevant edge layer.
At the transition, a reduction in the plasma turbulence and associated turbulent flux has been measured at the plasma edge region where the $E\times B$ flow shear is maximum, but also, though to a lesser extent, in the plasma core.
The reduction in the turbulence at the core, where the $E\times B$ flow shear is not substantially high, may be explained as a combined effect of the strong $E\times B$ flow shear at the edge that brakes the radial spreading of turbulence towards the core. Besides changes in the cross-phase between the density and the potential fluctuations may also contribute to the reduction of the turbulent flux at the core. 

New experiments are on-going at TJ-II aiming for a full characterization of turbulence and flows, including the impact of $E_r$ on cross-phases, density and temperature fluctuations. To that end, the already described diagnostics set is being upgraded and a new Gas Puff Imaging system is presently under commissioning at TJ-II. 
These prospective studies will hopefully provide a better understanding of the physical mechanism that selectively reduces only one of the transport channels, relevant not only for L-H transitions in stellarators but also for the I-mode in tokamaks. 
Besides, the role of fast particle driven modes on zonal flows and plasma turbulence will be also tackled to address the question on whether zonal flows can be directly driven by fast particle effects.

\ack{The authors acknowledge the entire TJ-II team for their support during the experiments. This work has been funded by the Spanish Ministry of Science and Innovation under contract numbers: ENE2007-65727, ENE2010-18409, ENE2013-48109-P, FIS2017-88892-P.
\\}

\bibliographystyle{prsty_copia}

\bibliography{Bibtex_database_copia}

\begin{thebibliography}{10}

\bibitem{Wagner:1982}
F. Wagner {\it et~al.}, Regime of Improved Confinement and High Beta in
  Neutral-Beam-Heated Divertor Discharges of the ASDEX Tokamak. Phys. Rev.
  Lett. {\bf 49},  1408  (1982).

\bibitem{Ryter:2002}
F. Ryter and the H-mode Threshold Database~Group, Progress of the international
  H-mode power threshold database activity. Plasma Phys. Control. Fusion {\bf
  44},  A415  (2002).

\bibitem{Wagner:2007}
F. Wagner, A quarter-century of H-mode studies. Plasma Phys. Control. Fusion
  {\bf 49},  B1  (2007).

\bibitem{Martin:2008}
Y.~R. Martin, T. Takizuka, and the ITPA CDBM H-mode Threshold Data~Group, Power
  requirement for accessing the H-mode in {ITER}. Journal of Physics:
  Conference Series {\bf 123},  012033  (2008).

\bibitem{Wagner:2006}
F. Wagner {\it et~al.}, H-mode and transport barriers in helical systems.
  Plasma Phys. Control. Fusion {\bf 48},  A217  (2006).

\bibitem{Itoh:1988}
S.~I. Itoh and K. Itoh, Model of L to H-Mode Transition in Tokamak. Phys. Rev.
  Lett. {\bf 60},  2276  (1988).

\bibitem{Biglari:1990}
H. Biglari, P. Diamond, and P. Terry, Influence of sheared polidal rotation on
  edge turbulence. Phys. Fluids B {\bf 2},  1  (1990).

\bibitem{Carreras:1994}
B.~A. Carreras, D. Newman, P.~H. Diamond, and Y.-M. Liang, Dynamics of low to
  high (L-H) confinement bifurcation: Poloidal flow and ion pressure gradient
  evolution. Physics of Plasmas {\bf 1},  4014  (1994).

\bibitem{Connor:2000}
J.~W. Connor and H.~R. Wilson, A review of theories of the L-H transition.
  Plasma Phys. Control. Fusion {\bf 42},  R1  (2000).

\bibitem{Kim:2003b}
E.-J. Kim and P.~H. Diamond, Zonal Flows and Transient Dynamics of the L-H
  Transition. Phys. Rev. Lett. {\bf 90},  185006  (2003).

\bibitem{Malkov:2009}
M.~A. Malkov and P.~H. Diamond, Weak hysteresis in a simplified model of the
  L-H transition. Phys. Plasmas {\bf 16},  012504  (2009).

\bibitem{Miki:2012}
K. Miki {\it et~al.}, Spatio-temporal evolution of the L-I-H transition. Phys.
  Plasmas {\bf 19},  092306  (2012).

\bibitem{Melnikov:2013}
A. Melnikov {\it et~al.}, Changes in plasma potential and turbulent particle
  flux in the core plasma measured by heavy ion beam probe during
  L{\textendash}H transitions in the {TJ}-{II} stellarator. Nuclear Fusion {\bf
  53},  092002  (2013).

\bibitem{Fernandez:2001}
A. Fern{\'a}ndez, W. Kasparek, K. Likin, and R. Mart{\'\i}n, Design of the
  Upgraded TJ-II Quasi-optical Transmission Line. International Journal of
  Infrared and Millimeter Waves {\bf 22},  649  (2001).

\bibitem{Liniers:2017}
M. Liniers {\it et~al.}, Beam transmission dependence on beam parameters for
  TJ-II Neutral Beam Injectors. Fusion Engineering and Design {\bf 123},  259
  (2017).

\bibitem{Pedrosa:2008}
M. Pedrosa {\it et~al.}, Evidence of Long-Distance Correlation of Fluctuations
  during Edge Transitions to Improved-Confinement Regimes in the {TJ-II}
  Stellarator. Phys. Rev. Lett. {\bf 100},  215003  (2008).

\bibitem{Hidalgo:2009}
C. Hidalgo {\it et~al.}, Multi-scale physics mechanisms and spontaneous edge
  transport bifurcations in fusion plasmas. EPL (Europhysics Letters) {\bf 87},
   55002  (2009).

\bibitem{Melnikov:2007}
A.~V. Melnikov {\it et~al.}, Plasma Potential Evolution Study by HIBP
  Diagnostic During NBI Experiments in the TJ-II Stellarator. Fusion Science
  and Technology {\bf 51},  31  (2007).

\bibitem{Melnikov:2015}
A. Melnikov {\it et~al.}, Control and data acquisition for dual HIBP
  diagnostics in the TJ-II stellarator. Fusion Engineering and Design {\bf
  96-97},  724  (2015), proceedings of the 28th Symposium On Fusion Technology
  (SOFT-28).

\bibitem{Happel:2009}
T. Happel {\it et~al.}, Doppler reflectometer system in the stellarator TJ-II.
  Rev. Sci. Instrum. {\bf 80},  073502  (2009).

\bibitem{Fujisawa:2004}
A. Fujisawa {\it et~al.}, Identification of Zonal Flows in a Toroidal Plasma.
  Phys. Rev. Lett. {\bf 93},  165002  (2004).

\bibitem{Milligen:1995b}
B. van Milligen {\it et~al.}, Wavelet bicoherence: A new turbulence analysis
  tool. Phys. Plasmas {\bf 2},  3017  (1995).

\bibitem{Grenfell:2020}
G. Grenfell {\it et~al.}, The impact of edge radial electric fields on
  edge{\textendash}scrape-off layer coupling in the {TJ}-{II} stellarator.
  Nuclear Fusion {\bf 60},  014001  (2020).

\bibitem{Sanchez:2009}
J. S{\'a}nchez {\it et~al.}, Confinement transitions in TJ-II under Li-coated
  wall conditions. Nuclear Fusion {\bf 49},  104018  (2009).

\bibitem{Tabares:2008}
F.~L. Tabar{\'{e}}s {\it et~al.}, Plasma performance and confinement in the
  {TJ}-{II} stellarator with lithium-coated walls. Plasma Physics and
  Controlled Fusion {\bf 50},  124051  (2008).

\bibitem{Estrada:2009}
T. Estrada {\it et~al.}, Sheared flows and transition to improved confinement
  regime in the TJ-II stellarator. Plasma Phys. Control. Fusion {\bf 51},
  124015  (2009).

\bibitem{Sano:2005}
F. Sano {\it et~al.}, H-mode confinement of Heliotron J. Nuclear Fusion {\bf
  45},  1557  (2005).

\bibitem{Hirsch:2008}
M. Hirsch {\it et~al.}, Major results from the stellarator Wendelstein 7-AS.
  Plasma Physics and Controlled Fusion {\bf 50},  053001  (2008).

\bibitem{Hubbard:2011}
A.~E. Hubbard {\it et~al.}, Edge energy transport barrier and turbulence in the
  I-mode regime on Alcator C-Mod. Physics of Plasmas {\bf 18},  056115  (2011).

\bibitem{Ryter:2016}
F. Ryter {\it et~al.}, I-mode studies at {ASDEX} Upgrade: L-I and I-H
  transitions, pedestal and confinement properties. Nuclear Fusion {\bf 57},
  016004  (2016).

\bibitem{Happel:2016}
T. Happel {\it et~al.}, Turbulence intermittency linked to the weakly coherent
  mode in {ASDEX} Upgrade I-mode plasmas. Nuclear Fusion {\bf 56},  064004
  (2016).

\bibitem{Hubbard:2016}
A. Hubbard {\it et~al.}, Multi-device studies of pedestal physics and
  confinement in the I-mode regime. Nuclear Fusion {\bf 56},  086003  (2016).

\bibitem{Hidalgo:2016}
C. Hidalgo, J. Talmadge, and M. Ramisch, Advancing the understanding of plasma
  transport in mid-size stellarators. Plasma Physics and Controlled Fusion {\bf
  59},  014051  (2016).

\bibitem{Manz:2021}
P. Manz {\it et~al.}, Gyrofluid simulation of an I-mode pedestal relaxation
  event. Physics of Plasmas {\bf 28},  102502  (2021).

\bibitem{Hirsch:2001}
M. Hirsch {\it et~al.}, Doppler reflectometry for the investigation of
  propagating density perturbations. Plasma Phys. Control. Fusion {\bf 43},
  1641  (2001).

\bibitem{Andrew:2008}
Y. Andrew {\it et~al.}, Evolution of the radial electric field in a {JET}
  H-mode plasma. {EPL} (Europhysics Letters) {\bf 83},  15003  (2008).

\bibitem{Silva:2021}
C. Silva {\it et~al.}, Structure of the {JET} edge radial electric field in He
  and D plasmas. Nuclear Fusion {\bf 61},  126006  (2021).

\bibitem{Milligen:2013}
B. van Milligen {\it et~al.}, Spatiotemporal and wavenumber resolved
  bicoherence at the low to high confinement transition in the {TJ}-{II}
  stellarator. Nuclear Fusion {\bf 53},  113034  (2013).

\bibitem{Diamond:2000}
P. Diamond {\it et~al.}, In search of the elusive zonal flow using
  cross-bicoherence analysis. Phys. Rev. Lett. {\bf 84},  4842  (2000).

\bibitem{Kobayashi:2017}
T. Kobayashi {\it et~al.}, Turbulent transport reduction induced by transition
  on radial electric field shear and curvature through amplitude and
  cross-phase in torus plasma. Scientific Reports {\bf 7},  14971  (2017).

\bibitem{Milligen:2021}
B.~P. van Milligen {\it et~al.}, Causality, intermittence, and crossphase
  evolution during confinement transitions in the TJ-II stellarator. Physics of
  Plasmas {\bf 28},  092302  (2021).

\bibitem{Kobayashi:2020}
T. Kobayashi, The physics of the mean and oscillating radial electric field in
  the L{\textendash}H transition: the driving nature and turbulent transport
  suppression mechanism. Nuclear Fusion {\bf 60},  095001  (2020).

\bibitem{Estrada:2011}
T. Estrada, C. Hidalgo, and T. Happel, Signatures of turbulence spreading
  during the H-L back-transition in TJ-II plasmas. Nuclear Fusion {\bf 51},
  032001  (2011).

\bibitem{Wang:2007}
W.~X. Wang {\it et~al.}, Nonlocal properties of gyrokinetic turbulence and the
  role of ExB flow shear. Phys. Plasmas {\bf 14},  072306  (2007).

\bibitem{Righi:1999}
E. Righi {\it et~al.}, Isotope scaling of the H mode power threshold on {JET}.
  Nuclear Fusion {\bf 39},  309  (1999).

\bibitem{Ryter:2013}
F. Ryter {\it et~al.}, Survey of the H-mode power threshold and transition
  physics studies in {ASDEX} Upgrade. Nuclear Fusion {\bf 53},  113003  (2013).

\bibitem{Losada:2018}
U. Losada {\it et~al.}, Role of isotope mass and evidence of fluctuating zonal
  flows during the L--H transition in the TJ-II stellarator. Plasma Physics and
  Controlled Fusion {\bf 60},  074002  (2018).

\bibitem{Losada:2021}
U. Losada {\it et~al.}, Spatial characterization of edge zonal flows in the
  {TJ}-{II} stellarator: the roles of plasma heating and isotope mass. Plasma
  Physics and Controlled Fusion {\bf 63},  044002  (2021).

\bibitem{Estrada:2010c}
T. Estrada {\it et~al.}, Experimental observation of coupling between
  turbulence and sheared flows during L-H transitions in a toroidal plasma.
  Europhysics Letters {\bf 92},  35001  (2010).

\bibitem{Estrada:2011b}
T. Estrada, C. Hidalgo, T. Happel, and P.~H. Diamond, Spatiotemporal Structure
  of the Interaction between Turbulence and Flows at the L-H Transition in a
  Toroidal Plasma. Phys. Rev. Lett. {\bf 107},  245004  (2011).

\bibitem{Miki:2013}
K. Miki {\it et~al.}, Spatio-temporal evolution of the L - H and H - L
  transitions. Nuclear Fusion {\bf 53},  073044  (2013).

\bibitem{Xu_GS:2011}
G.~S. Xu {\it et~al.}, First Evidence of the Role of Zonal Flows for the L-H
  Transition at Marginal Input Power in the EAST Tokamak. Phys. Rev. Lett. {\bf
  107},  125001  (2011).

\bibitem{Schmitz:2012}
L. Schmitz {\it et~al.}, Role of Zonal Flow Predator-Prey Oscillations in
  Triggering the Transition to H-Mode Confinement. Phys. Rev. Lett. {\bf 108},
  155002  (2012).

\bibitem{Kobayashi:2013}
T. Kobayashi {\it et~al.}, Spatiotemporal Structures of Edge Limit-Cycle
  Oscillation before L-to-H Transition in the JFT-2M Tokamak. Phys. Rev. Lett.
  {\bf 111},  035002  (2013).

\bibitem{Tynan:2013}
G. Tynan {\it et~al.}, Turbulent-driven low-frequency sheared $E\times B$ flows
  as the trigger for the H-mode transition. Nuclear Fusion {\bf 53},  073053
  (2013).

\bibitem{Cheng:2013}
J. Cheng {\it et~al.}, Dynamics of Low-Intermediate-High-Confinement
  Transitions in Toroidal Plasmas. Phys. Rev. Lett. {\bf 110},  265002  (2013).

\bibitem{Estrada:2015}
T. Estrada {\it et~al.}, Limit cycle oscillations at the L--I--H transition in
  TJ-II plasmas: triggering, temporal ordering and radial propagation. Nuclear
  Fusion {\bf 55},  063005  (2015).

\bibitem{Happel:2011}
T. Happel {\it et~al.}, Scale-selective turbulence reduction in H-mode plasmas
  in the TJ-II stellarator. Phys. Plasmas {\bf 18},  102302  (2011).

\bibitem{Estrada:2012b}
T. Estrada {\it et~al.}, Spatial, temporal and spectral structure of the
  turbulence-flow interaction at the L-H transition. Plasma Phys. Control.
  Fusion {\bf 54},  124024  (2012).

\bibitem{Manz:2009}
P. Manz, M. Ramisch, and U. Stroth, Physical Mechanism behind Zonal-Flow
  Generation in Drift-Wave Turbulence. Phys. Rev. Lett. {\bf 103},  165004
  (2009).

\bibitem{Stroth:2011}
U. Stroth, P. Manz, and M. Ramisch, On the interaction of turbulence and flows
  in toroidal plasmas. Plasma Physics and Controlled Fusion {\bf 53},  024006
  (2011).

\bibitem{Scott:2005}
B.~D. Scott, Energetics of the interaction between electromagnetic ExB
  turbulence and zonal flows. New Journal of Physics {\bf 7},  92  (2005).

\bibitem{Alonso:2012}
J. Alonso {\it et~al.}, Dynamic transport regulation by zonal flow-like
  structures in the {TJ}-{II} stellarator. Nuclear Fusion {\bf 52},  063010
  (2012).

\bibitem{Gerru:2019}
R. Gerr{\'{u}} {\it et~al.}, On the interplay between turbulent forces and
  neoclassical particle losses in zonal flow dynamics. Nuclear Fusion {\bf 59},
   106054  (2019).

\bibitem{Wobig:2000}
H. Wobig and J. Kisslinger, Viscous damping of rotation in Wendelstein 7-AS.
  Plasma Phys. Control. Fusion {\bf 42},  823  (2000).

\bibitem{Estrada:2010a}
T. Estrada {\it et~al.}, L-H Transition Experiments in TJ-II. Contrib. Plasma
  Phys. {\bf 50},  501  (2010).

\bibitem{Bondarenko:2010}
O. Bondarenko {\it et~al.}, Influence of low order rational surfaces on the
  radial electric field of TJ-II ECH plasmas. Contrib. Plasma Phys. {\bf 50},
  605  (2010).

\bibitem{Estrada:2016}
T. Estrada {\it et~al.}, Plasma flow, turbulence and magnetic islands in TJ-II.
  Nuclear Fusion {\bf 56},  026011  (2016).

\bibitem{Ida:2001}
K. Ida {\it et~al.}, Observation of Plasma Flow at the Magnetic Island in the
  Large Helical Device. Phys. Rev. Lett. {\bf 88},  015002  (2001).

\bibitem{Zhao:2015}
K. Zhao {\it et~al.}, Plasma flows and fluctuations with magnetic islands in
  the edge plasmas of J-{TEXT} tokamak. Nuclear Fusion {\bf 55},  073022
  (2015).

\bibitem{Bardoczi:2016}
L. Bard\'oczi {\it et~al.}, Modulation of Core Turbulent Density Fluctuations
  by Large-Scale Neoclassical Tearing Mode Islands in the DIII-D Tokamak. Phys.
  Rev. Lett. {\bf 116},  215001  (2016).

\bibitem{Choi:2017}
M. Choi {\it et~al.}, Multiscale interaction between a large scale magnetic
  island and small scale turbulence. Nuclear Fusion {\bf 57},  126058  (2017).

\bibitem{Estrada:2021b}
T. Estrada {\it et~al.}, Impact of magnetic islands on plasma flow and
  turbulence in W7-X. Nuclear Fusion {\bf 61},  096011  (2021).

\bibitem{Banon_Navarro:2017}
A. Ba{\~{n}}{\'{o}}n-Navarro {\it et~al.}, Effect of magnetic islands on
  profiles, flows, turbulence and transport in nonlinear gyrokinetic
  simulations. Plasma Phys. Control. Fusion {\bf 59},  034004  (2017).

\bibitem{DiSiena:2020}
A.~D. Siena, A. Ba{\~{n}}{\'{o}}n-Navarro, and F. Jenko, Turbulence Suppression
  by Energetic Particle Effects in Modern Optimized Stellarators. Phys. Rev.
  Lett. {\bf 125},  105002  (2020).

\bibitem{Chen:2012}
L. Chen and F. Zonca, Nonlinear Excitations of Zonal Structures by Toroidal
  Alfv\'en Eigenmodes. Phys. Rev. Lett. {\bf 109},  145002  (2012).

\bibitem{Qiu:2019}
Z. Qiu, L. Chen, F. Zonca, and W. Chen, Nonlinear excitation of a geodesic
  acoustic mode by toroidal Alfv{\'{e}}n eigenmodes and the impact on plasma
  performance. Nuclear Fusion {\bf 59},  066031  (2019).

\bibitem{Mazzi:2020}
S. Mazzi {\it et~al.}, Impact of fast ions on a trapped-electron-mode dominated
  plasma in a {JT}-60U hybrid scenario. Nuclear Fusion {\bf 60},  046026
  (2020).

\bibitem{Jimenez:2007}
R. Jim{\'e}nez-G{\'o}mez {\it et~al.}, Analysis of magnetohydrodynamic
  instabilities in {TJ-II} plasmas. Fusion Science and Technology {\bf 51},  20
   (2007).

\bibitem{Hidalgo:2022}
C. Hidalgo {\it et~al.}, Overview of the TJ-II stellarator research programme
  towards model validation in fusion plasmas. Nucl. Fusion {\bf 62},    (2022).

\bibitem{Schmitz:2017}
L. Schmitz, The role of turbulence{\textendash}flow interactions in L- to
  H-mode transition dynamics: recent progress. Nuclear Fusion {\bf 57},  025003
   (2017).

\bibitem{Bourdelle:2020}
C. Bourdelle, Staged approach towards physics-based L-H transition models.
  Nuclear Fusion {\bf 60},  102002  (2020).

\end{thebibliography}

\end{document}